\documentclass[conference]{IEEEtran}
\IEEEoverridecommandlockouts
\usepackage{cite}
\usepackage{amsmath,amssymb,amsfonts}
\usepackage{algorithmic}
\usepackage{graphicx}
\usepackage{textcomp}
\usepackage{xcolor}
\usepackage[caption=false,font=footnotesize]{subfig}
\usepackage{mathtools}
\usepackage{makecell}

\usepackage{amssymb} 
\usepackage{amsmath,amsfonts,bm,stmaryrd}
\usepackage[thinc]{esdiff} 

\usepackage{xcolor}

\usepackage[mode=image]{standalone}

\usepackage{pgfplots}
\pgfplotsset{compat=newest}

\usepackage[outline]{contour}

\usepackage{anyfontsize}

\usepackage[short,nocomma]{optidef}

\usepackage{cite} 

\usetikzlibrary{fadings,shapes.geometric,positioning,calc,arrows.meta,backgrounds,bending,arrows,decorations.markings,decorations.pathreplacing,intersections,fit,spy}\usetikzlibrary{arrows,shapes,calc,positioning,decorations.markings,intersections,scopes,matrix,fit,chains}
\usepackage[edges]{forest}

\newlength\fheight
\newlength\fwidth 

\def\BibTeX{{\rm B\kern-.05em{\sc i\kern-.025em b}\kern-.08em
    T\kern-.1667em\lower.7ex\hbox{E}\kern-.125emX}}
\begin{document}

\title{Distortions Characterization for Dynamic Carrier Allocation in Ultra High-Throughput Satellites} 

\author{\IEEEauthorblockN{Tony Colin, Thomas Delamotte and Andreas Knopp}
\IEEEauthorblockA{\textit{Bundeswehr University Munich} \\
\textit{Chair of Signal Processing}\\
Neubiberg, Germany \\
\{paper.sp, tony.colin\}@unibw.de}
%
%
%
}

\maketitle

\begin{abstract}
A novel analytical formula for the characterization of linear and nonlinear distortions in future ultra high-throughput communication payloads is proposed in this work. 
In this context, the carrier-to-interference ratio related to single-carrier and multicarrier signals is derived. Through the analysis of its behavior valuable insights are created, especially regarding the interaction between linear and nonlinear intersymbol interference. 
Furthermore, the principle of dynamic carrier allocation optimization is highlighted in a realistic scenario. Within the presented framework, it is proven that a significant gain can be achieved even with a limited number of carriers. 
Finally, a complexity and accuracy analysis emphasizes the practicality of the proposed approach. 
%
%
\end{abstract}


\section{Introduction}
Ultra high-throughput satellite (UHTS) systems development has been driven by the ever-increasing demand for high data rate satellite services.
For the next generation of UHTS systems, it is foreseen that up to 2.9 GHz of downlink bandwidth per polarization will be achieved by shifting feeder links to the Q/V-band and exploiting the available Ka-band resources in the user links \cite{Vidal}. 
This will help to meet the need of high data rate services for future beyond 5G communications (e.g., broadband internet services and delivery to the edge) \cite{Gaudenzi}.
Furthermore, the allocation of payload resources (e.g., bandwidth, time, power, and coverage) will become increasingly flexible, 
especially through the evolution of digital processors. 
This will enable to adapt to different communication services and traffic demands.
%

%
In particular, a dynamic carrier allocation (DCA) within the satellite will pave the way to a fully flexible connectivity between uplink and downlink beams. 
On one side, this strategy removes the need of a centralized gateway. One the other side, it prevents from increasing the processing power onboard the satellite, which is a major drawback of alternative strategies such as beam-hopping (physical layer) or flexible routing protocols (higher layer). 
%
%
%
%
%
The DCA is well known in the context of terrestrial communication systems, where high capacity gains can be achieved \cite{DCAMobile}. However, it is still an uncharted territory in the context of flexible satellite payloads. 
%
Indeed, conventional approaches rely on static carrier allocation, which under-exploits the available frequency resources and does not allow any adaptation to the highly varying traffic demand. 
To enable the DCA in UHTS systems, two main phenomena need to be taken into account: the linear distortions and the nonlinear distortions. 
On the one hand, the analytical characterization and modeling of the nonlinearities  entailed by the satellite high power amplifier (HPA) have been well covered in the literature \cite{Saleh2}, \cite{Beidas2011IMD}.
On the other hand, the work of \cite{Colin} has recently enabled the analytical characterization of the linear distortions in UHTS systems, essentially entailed by the wideband output multiplexer (OMUX) filters. 
Therefore, it is currently possible to describe efficiently the main frequency-dependent components in the link budget and optimize the allocation of each user carrier. 
This key technology will enable to further increase the capacity of the next-generation UHTS systems. 

Thus, the original contribution of the present paper with respect to the state of the art is the following:
\begin{itemize}
\item A novel analytical characterization of both linear and nonlinear intersymbol interference is derived, enabling the study of their mutual interaction. The results are validated by simulations.
\item A low-complexity analytical formula of the carrier-to-interference ratio (CIR) is derived in order to be exploited in the context of DCA optimization.
\item The feasibility of near real-time DCA is emphasized in a practical scenario. Within this framework, the accuracy and complexity of the proposed CIR formula, along with the allocation gain are discussed.
\end{itemize}

The remainder of this paper is structured as follows. Section II covers the UHTS system model. In Section III, the linear and nonlinear distortions are thoroughly analyzed and the total CIR formula is derived. Section IV contains a practical DCA scenario. The conclusions are presented in Section V.

\begin{figure*}[!t]
	\centering 
	\includegraphics[scale=1.34]{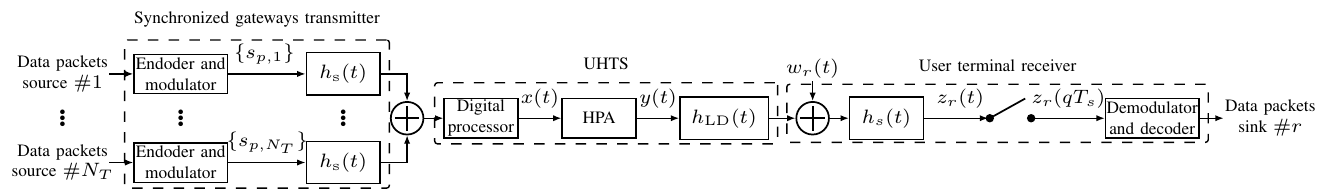}\\
	\vspace{-0.1cm}
	\caption{Block diagram of the future UHTS communication system.} 
	\label{fig:Transmission-BD} 
	\vspace{-0.3cm}
\end{figure*}
%
\section{UHTS System Model}
The UHTS system model is depicted in Fig.~\ref{fig:Transmission-BD}. The forward link of a geostationary UHTS is considered. This means that the uplink white Gaussian noise (WGN) can be neglected due to better link budget in the feeder link. The downlink WGN at the $r^{\mathrm{th}}$-user receiver terminal has a power of $\sigma_r^2$ and is denoted $w_r(t)$. Furthermore, the commonly used square-root raised cosine filter (SRRCF) $h_s(t)$ is considered as pulse shaping and matched filter. The roll-off factor is denoted $\alpha$. 

Thus, the single-carrier transmit signal ($N_T=1$)\footnote{The extension to multicarrier signals is performed in the next section.} after pulse shaping, used as a basis for the later calculations, can be expressed as follows:
\begin{equation} \label{sendsign}
x(t)=\sum_{p=0}^{N_s-1} s_{p}\cdot h_s(u-p T_s)\ ,
\end{equation}
where the realizations $\{s_p\}_{0\leq p\leq N_s-1}$ are associated to the random variables $\{S_p\}_{0\leq p\leq N_s-1}$ considered uncorrelated. 
The number of symbols and the symbol duration are denoted $N_s$ and $T_s$, respectively. 
The symbols are sent with a fixed average transmit power $\mathrm{V}[S_p]=P$, $\forall p\in \{0;N_s-1\}$. $E[\mathcal{Z}]$ and $\mathrm{V}[\mathcal{Z}]$ are the expectation and variance of a complex variable $\mathcal{Z}$, respectively, where $\mathrm{V}[\mathcal{Z}]=E[|\mathcal{Z}|^2]-|E[\mathcal{Z}]|^2$. 
For the sake of simplicity, it is assumed in the next section that the uplink gains and losses have been taken into account and that $P$ corresponds to the symbols power at the input of the HPA. 

\section{Distortions Characterization} 
In this section, an analytical expression of the CIR is determined. 
Firstly, the nonlinear intersymbol interference (NL-ISI) and linear intersymbol interference (L-ISI) are studied for a single-carrier signal. Then, an extension to multicarrier signals and adjacent channel interference (ACI) is discussed.
%
\subsection{Single-Carrier Analysis} 
\subsubsection{Nonlinear Distortions}
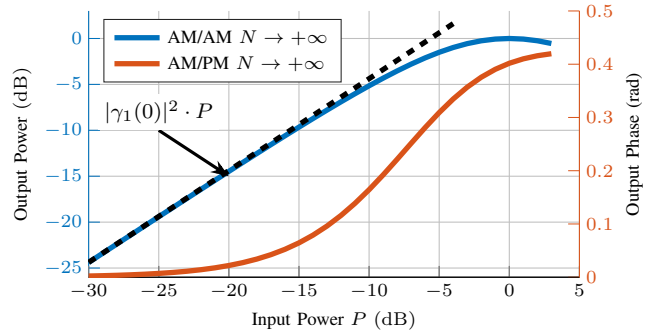
\begin{figure}[b!]
	\centering 
	\scriptsize 
	\vspace{-0.5cm}
	\setlength\fheight{0.4\columnwidth} 
	\setlength\fwidth{0.34\columnwidth}
%
%
\definecolor{mycolor1}{rgb}{0.00000,0.44700,0.74100}%
\definecolor{mycolor2}{rgb}{0.85000,0.32500,0.09800}%
\begin{tikzpicture}

\begin{axis}[%
width=2.1667\fwidth,
height=\fheight,
at={(0\fwidth,0\fheight)},
scale only axis,
xmin=-30,
xmax=5,
xticklabels={},
xlabel style={font=\color{white!15!black}},
xlabel={}, 
every outer y axis line/.append style={mycolor1},
every y tick label/.append style={font=\color{mycolor1}},
every y tick/.append style={mycolor1},
ymin=-26,
ymax=3,
ytick={-25,-20,-15,-10,-5,0},
ylabel={Output Power$\ \mathrm{(dB)}$},
title style={font=\bfseries, align=center},
title={}, 
axis x line*=bottom,
axis y line*=left,
xmajorgrids,
ymajorgrids,
legend style={at={(0.03,0.97)}, anchor=north west, legend cell align=left, align=left, draw=white!15!black}
]
\addplot [color=mycolor1, line width=2.0pt] 
  table[row sep=crcr]{%
-30	-24.3910778394583\\
-29	-23.3934788602751\\
-28	-22.3964964324268\\
-27	-21.4002872424357\\
-26	-20.4050468666533\\
-25	-19.4110189044722\\
-24	-18.4185059734464\\
-23	-17.4278827966568\\
-22	-16.4396115640818\\
-21	-15.4542596596476\\
-20	-14.4725197142257\\
-19	-13.4952317980484\\
-18	-12.523407485551\\
-17	-11.5582556967897\\
-16	-10.6012109918413\\
-15	-9.65396692437953\\
-14	-8.71852084708485\\
-13	-7.79724269351029\\
-12	-6.89298811733316\\
-11	-6.00928264894226\\
-10	-5.15060060913244\\
-9	-4.32273893183068\\
-8	-3.53323273627566\\
-7	-2.79168343775547\\
-6	-2.10980742015744\\
-5	-1.50102135266166\\
-4	-0.979496938924088\\
-3	-0.558811805992132\\
-2	-0.250493662329744\\
-1	-0.0628023788733515\\
0	-9.64327466553287e-16\\
1	-0.062189783648515\\
2	-0.245653441274106\\
3	-0.543532093703503\\
};
\addlegendentry{AM/AM $N\rightarrow+\infty$}


\addplot [color=mycolor2, line width=2.0pt] 
  table[row sep=crcr]{%
-30	10\\
};
\addlegendentry{AM/PM $N\rightarrow+\infty$}

\addplot [color=black, dashed, line width=2.0pt]
  table[row sep=crcr]{%
-30	-24.3817705931989\\
-4	1.6182294068011\\
};

\draw [black, arrows={-Stealth[scale=0.8]}, line width=1.2pt,xshift=0cm,yshift=0cm] (-25,-10) -- (-20,-15);

\draw[draw=none,fill=white,opacity=1.0] (-25,-8) node [font=\footnotesize,fill=white,opacity=1.0] {{\color{black}$|\gamma_1(0)|^2\cdot P$}};

\end{axis}

\begin{axis}[%
width=2.1667\fwidth,
height=\fheight,
at={(0\fwidth,0\fheight)},
scale only axis,
xmin=-30,
xmax=5,
xlabel={Input Power $P\ \mathrm{(dB)}$}, 
every outer y axis line/.append style={mycolor2},
every y tick label/.append style={font=\color{mycolor2}},
every y tick/.append style={mycolor2},
ymin=0,
ymax=0.5,
ytick={0,0.1,0.2,0.3,0.4,0.5},
ylabel={Output Phase (rad)},
title style={font=\bfseries, align=center},
title={}, 
axis x line*=bottom,
axis y line*=right,
legend style={at={(0.03,0.97)}, anchor=north west, legend cell align=left, align=left, draw=white!15!black}
]
\addplot [color=mycolor2, line width=2.0pt] 
  table[row sep=crcr]{%
-30	0.00227006925703677\\
-29	0.00285519358861919\\
-28	0.00359027558223346\\
-27	0.00451324542859766\\
-26	0.0056713358381657\\
-25	0.00712319377800341\\
-24	0.00894137160828024\\
-23	0.011215204099873\\
-22	0.0140540317764675\\
-21	0.0175906508239692\\
-20	0.0219847388405014\\
-19	0.027425803139792\\
-18	0.0341349017912265\\
-17	0.0423639810386938\\
-16	0.0523911641466306\\
-15	0.0645097813312212\\
-14	0.0790085237203009\\
-13	0.0961401869174452\\
-12	0.116077605377399\\
-11	0.138858257153789\\
-10	0.164324087863985\\
-9	0.192069815856261\\
-8	0.221418901666187\\
-7	0.251447074301439\\
-6	0.281064425029452\\
-5	0.30914841933815\\
-4	0.334698851967875\\
-3	0.356973588356201\\
-2	0.37556905507066\\
-1	0.390429544185082\\
0	0.401793465610023\\
1	0.41010134862163\\
2	0.415894682225137\\
3	0.419728631300963\\
};


\end{axis}

\end{tikzpicture}%
	\caption{Saleh TWTA characteristic (normalized parameters for typical TWTA).}
	\label{fig:Saleh} 
\end{figure}

The most widespread memoryless HPA model is the Saleh model \cite{Saleh1}. This model describes mathematically the HPA nonlinearities from an input signal $x(t)$ to an output signal $y(t)$ depending on $|x(t)|^2$, the input signal instantaneous power at time $t$. In this paper, the normalized parameters for a typical HPA are assumed as described in \cite{Saleh2}. The corresponding AM/AM and AM/PM characteristics are depicted in Fig.~\ref{fig:Saleh}. Moreover, the input power $P$ is defined such that it corresponds to the input back-off (the input and output power at saturation are normalized). 
%
%
%
%

To improve the mathematical tractability, it is convenient to consider the truncated memoryless polynomial:
\begin{equation} \label{HPApoly}
y(t)=x(t)\cdot \sum_{k=0}^{N} \gamma_{2k+1}(a) \cdot |x(t)|^{2k}\ ,
\end{equation}
where $\{\gamma_{2k+1}(a)\}_{0\leq k\leq N}$ are the complex coefficients resulting from the Taylor approximation of the Saleh model when the instantaneous power $|x(t)|^2$ gets close to a constant $a$. 
$N$ refers to to the number of odd order distortions to be considered. 

Without loss of generality, the case $N=1$ is considered. Thus, the relevant polynomial coefficients are:
\begin{equation} \label{salehpolycoeff}
\begin{aligned}[c] 
&\gamma_1(a)=\frac{\alpha_{\mathcal{I}} (1+2\beta_{\mathcal{I}} a)}{(1+\beta_{\mathcal{I}} a)^2}\ ,\\
&\gamma_3(a)=-\frac{\alpha_{\mathcal{I}} \beta_{\mathcal{I}}}{(1+\beta_{\mathcal{I}} a)^2}+j\cdot \frac{\alpha_{\mathcal{Q}}}{(1+\beta_{\mathcal{Q}} a)^2}\ ,
\end{aligned}
\end{equation}
where $\gamma_1(a)$ and $\gamma_3(a)$ are the HPA linear gain and the third order complex coefficient, respectively, when the input back-off gets close to the value $a$. Furthermore, $\alpha_{\mathcal{I}}=1.90947$, $\beta_{\mathcal{I}}=1.07469$, $\alpha_{\mathcal{Q}}=4.35023$, and $\beta_{\mathcal{Q}}=2.33525$ are the normalized Saleh parameters. The dependency on the variables $a$ and $r$ will be omitted in the following for more compact notations. 

Given the typical complex coefficients, the received signal can be approximated as:
\begin{equation} \label{recsmpl}
z(q T_s) = \gamma_1\cdot z_1(q T_s) + \gamma_3\cdot z_3(q T_s) + (w*h_s)(q T_s)\ ,
\end{equation}
where
%
\begin{equation} \label{recsmplcmp2}
\begin{aligned}[c] 
&z_1(q T_s)=\sum_{k_1\in\mathbb{Z}} s_{q-k_1}\cdot \int_{\mathbb{R}} h^\ddagger(u) h_s(u+k_1 T_s) du\ ,\\
&z_3(q T_s)=\sum_{k_1\in\mathbb{Z}} \sum_{k_2\in\mathbb{Z}} \sum_{k_3\in\mathbb{Z}} s_{q-k_1}s_{q-k_2}s^*_{q-k_3}\cdot \\
&\int_{\mathbb{R}} h^\ddagger(u) h_s(u+k_1 T_s) h_s(u+k_2 T_s) h_s(u+k_3 T_s) du\ ,
\end{aligned}
\end{equation}
are the received sampled signal components of first and third order, respectively. They are obtained after some mathematical manipulations similar to \cite{ICC98}. The filter border effects at the beginning and at the end of the symbol sequence are also neglected. $h^\ddagger(u)$ represents the post-HPA filters, which means that without linear distortions $h^\ddagger(u)=h_s(u)$.
%
\subsubsection{Linear Distortions}
\begin{figure}[b!]
	\centering 
	\scriptsize 
	\vspace{-0.5cm}
	\setlength\fheight{0.4\columnwidth} 
	\setlength\fwidth{0.34\columnwidth}
	\input{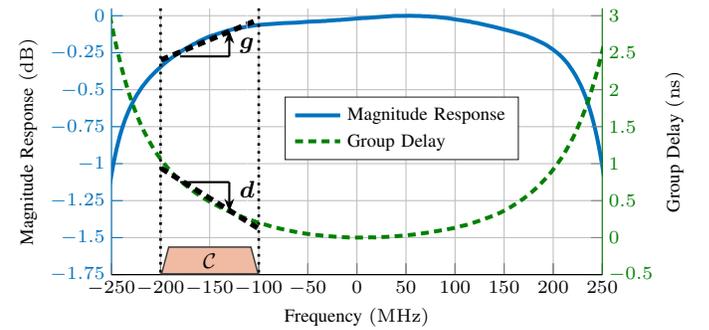}
	\vspace{-0.4cm}
	\caption{Re-scaled DVB-S2x OMUX filter characteristic including an example of gain slope $g$ and group delay slope $d$ approximation with regards to a carrier $\mathcal{C}$ of bandwidth $200\ \mathrm{MHz}$ and center frequency $-150\ \mathrm{MHz}$.}
	\label{fig:DVBS2xCharac} 
\end{figure}

Since the linear distortions caused by the OMUX filter after the HPA are dominant \cite{Colin}, the gain and group delay variations model will be based -- without loss of generality -- on the realistic re-scaled OMUX characteristic as illustrated in Fig.~\ref{fig:DVBS2xCharac}. 
%
A first-order approximation of the linear distortions has been proven to be very accurate (especially in a multicarrier configuration), while greatly lowering the implementation complexity \cite{Colin}. This so-called slope-based model relies on the gain slope $g$ $\mathrm{(dB/MHz)}$ and group delay slope $d$ $\mathrm{(ns/MHz)}$. These slopes are determined by least-squares approximations over the considered carrier bandwidth. 
Thus, the post-HPA impulse response can be expressed as:
\begin{equation} \label{impLD1}
h^\ddagger (u)=\int_{-\infty}^{+\infty} H_{\mathrm{LD}}(f)\sqrt{H_{\mathrm{RCF}}(f)}e^{j2\pi f u}df\ ,
\end{equation}
where
\begin{equation} \label{impLD2}
H_{\mathrm{LD}}(f) = e^{\xi g f - j\pi d f^2}\ ,
\end{equation}
is the first-order linear distortions transfer function and $\xi=\ln(10)/20$. It is trivial to see that the frequency-dependent gain and group delay variations are affecting all kernels through $h^\ddagger (u)$. As discussed in \cite{Colin}, it is more convenient to normalize the expressions with respect to the symbol rate $R_s$. Thus, the only two linear distortions variables are $x_g=g\cdot R_s$ and $y_d=d\cdot R_s^2$ and expressed in $\mathrm{(dB/MHz)(Mbauds)}$ and $\mathrm{(ns/MHz^2)(Mbauds)^2}$. 
%
\subsubsection{Carrier-to-Interference Ratio}
The analytical determination of the CIR\footnote{Since the sampled filtered noise power results in a constant independent of the frequency, it can be omitted from the analysis and allocation optimization.} can be viewed as the ability to separate the useful signal power (constructive) from the interference power (destructive) in the expression of $\mathrm{V}[z(q T_s)]$. 
The variance of the received sampled signal can be expressed as:
\begin{equation} \label{CIR_VarTot}
\begin{aligned}[c] 
\mathrm{V}[z(q T_s)]=&|\gamma_1|^2 \cdot \mathrm{V}[z_1(q T_s)] + |\gamma_3|^2 \cdot \mathrm{V}[z_3(q T_s)] +\\
&\gamma_1 \gamma_3^* \cdot \mathrm{Cov}[z_1(q T_s);z_3(q T_s)]+\\
&\gamma_3 \gamma_1^* \cdot \mathrm{Cov}[z_3(q T_s);z_1(q T_s)]+\\
&\mathrm{V}[(w*h_s)(q T_s)]\ .
\end{aligned}
\end{equation}
Thus, each frequency-dependent term can be divided into its useful and interfering part, such as:
%
\begin{equation} \label{CIR_VarA}
\begin{aligned}[c] 
\mathrm{V}[z_1(q T_s)]&=P\cdot (\kappa_{u,1,1} + \kappa_{i,1,3}),\\
\mathrm{V}[z_3(q T_s)]&=P^3\cdot (\kappa_{u,3,3} + \kappa_{i,3,3}),\\
\mathrm{Cov}[z_1(q T_s);z_3(q T_s)]&=P^2\cdot (\kappa_{u,1,3} + \kappa_{i,1,3}),\\
\mathrm{Cov}[z_3(q T_s);z_1(q T_s)]&=P^2\cdot (\kappa_{u,3,1} + \kappa_{i,3,1}),
\end{aligned}
\end{equation}
where, for instance,
\begin{equation} \label{CIR_VarB1}
\begin{aligned}[c] 
\kappa_{u,1,1}=&\left\lvert \int_{\mathbb{R}} h^\ddagger(u)h_s(u) du \right\rvert^2\ , \\
\kappa_{i,1,1}=&\sum_{k_1\in\mathbb{Z}\backslash \{0\}} \left\lvert \int_{\mathbb{R}} h^\ddagger(u) h_s(u+k_1 T_s) du \right\rvert^2\ ,
\end{aligned}
\end{equation}
are the useful and interfering part of the Volterra kernel power related to $z_1(qT_s)$. 
As opposed to the first order, the third order of nonlinear distortions contains many terms to separate. The useful and interference terms start as follows:
\begin{equation} \label{CIR_VarB3}
\begin{aligned}[c] 
\kappa_{u,3,3}=&\sum_{k_2,k_3\in\mathbb{Z}} \mathcal{H}(0,k_2,k_3)+\cdots\ , \\
\kappa_{i,3,3}=&\sum_{\substack{k_1\in\mathbb{Z}\backslash \{0\};k_2,k_3\in\mathbb{Z}}} \mathcal{H}(k_1,k_2,k_3)+\cdots\ ,
\end{aligned}
\end{equation}
where
\begin{equation} \label{func}
\begin{aligned}[c] 
&\mathcal{H}(k_1,k_2,k_3)=\\
&\left\lvert \int_{\mathbb{R}} h^\ddagger(u)h_s(u+k_1 T_s) h_s(u+k_2 T_s) h_s(u+k_3 T_s) du \right\rvert^2\ .\nonumber
\end{aligned}
\end{equation}
%
Thus, the single-carrier CIR can be derived based on this decomposition and can be expressed as:
%
%
%
\begin{equation} \label{CIR_D2}
\left(\frac{C_{\mathrm{ISI}}}{I_{\mathrm{ISI}}}\right)=
\frac{
P\cdot c_1
+ P^2\cdot c_2
+ P^3\cdot c_3
}
{
P\cdot i_1
+ P^2\cdot i_2
+ P^3\cdot i_3
}\ ,
\end{equation}
where
%
%
\begin{equation*} \label{CIR_VarDef}
\begin{split}
c_1&=|\gamma_1|^2\cdot \kappa_{u,1,1}\ ,\ \qquad\\
c_2&=\gamma_1\gamma_3^* \cdot \kappa_{u,1,3}\ \ \qquad\\
&+\gamma_1^*\gamma_3 \cdot \kappa_{u,3,1}\ ,\ \qquad\\
c_3&=|\gamma_3|^2\cdot \kappa_{u,3,3}\ ,\ \qquad\\
\end{split}
\begin{split}
i_1&=|\gamma_1|^2\cdot \kappa_{i,1,1}\ ,\\
i_2&=\gamma_1\gamma_3^* \cdot \kappa_{i,1,3}\\
&+\gamma_1^*\gamma_3 \cdot \kappa_{i,3,1}\ ,\\
i_3&=|\gamma_3|^2\cdot \kappa_{i,3,3}\ .\\
\end{split} \nonumber
\end{equation*}
%
%
It is worth noting that with non-constant modulus symbol constellations, e.g. in the case 16APSK symbols, an additional factor depending on the ring ratio comes out of the variance terms in $P^2$ and $P^3$. This factor is worsening the terms in $P^2$ and $P^3$ around the saturation. 

\begin{figure}[t!]
	\centering 
	\scriptsize 
	\setlength\fheight{0.6\columnwidth} 
	\setlength\fwidth{0.6\columnwidth}
	\input{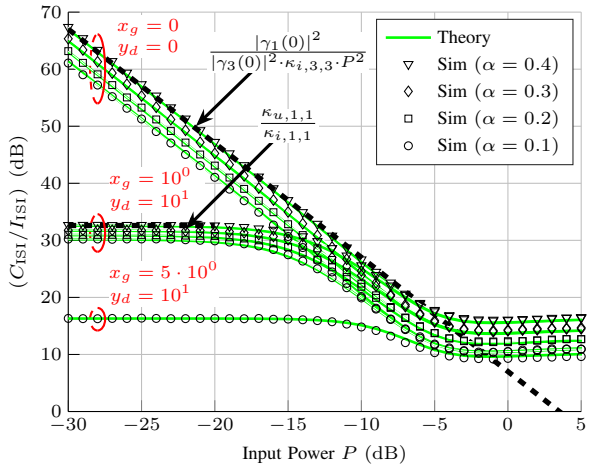}
	\vspace{-0.3cm}
	\caption{Single-carrier CIR analysis. The units of $x_g$ and $y_d$ are $\mathrm{(dB/MHz)\cdot (Mbauds)}$ and $\mathrm{(ns/MHz)\cdot (Mbauds)^2}$, respectively. The Saleh model coefficients are approximated at $a=0$ for $N=1$ and the sums limit\protect\footnotemark is $L=10$.}
	\vspace{-0.4cm}
	\label{fig:CID3} 
\end{figure}

Figure~\ref{fig:CID3} illustrates the novel characterization of the CIR related to ISI as a function of the input power $P$, the roll-off $\alpha$, and the amount of linear distortions represented by $x_g$ and $y_d$. 
To best describe the interactions between L-ISI and NL-ISI, the limit cases are first discussed. 
On the one hand, only NL-ISI are present in the system when $x_g=y_d=0$. This leads to $\kappa_{u,1,1}=1$ and $\kappa_{i,1,1}=\kappa_{i,1,3}=\kappa_{i,3,1}=0$ based on the properties of the SRRCF. In this case, it can be observed that the lower the roll-off, the lower the CIR. 
On the other hand, only L-ISI are present in the system when the HPA is operated in the linear region, i.e. $P\rightarrow 0$. This leads to $(C_{\mathrm{ISI}}/I_{\mathrm{ISI}})\rightarrow \kappa_{u,1,1}/\kappa_{i,1,1}$, which exactly corresponds to the closed-form solution derived in \cite{Colin}. 

Two examples of linear distortions are given to highlight the trade-off between L-ISI and NL-ISI. Both are considered in the case of a dominant gain slope, which happens for instance when a carrier is located at the edge of the wideband OMUX. 
When $x_g=10^0\ \mathrm{dB/MHz\cdot Mbauds}$, higher degree kernels are not affected by linear distortions and a transition region is observed around $P_{\mathrm{dB}}=-9\ \mathrm{dB}$ where the lower degree kernels become dominant. It is only when the linear distortions are stronger, i.e. $\kappa_{u,1,1}/\kappa_{i,1,1}$ lying below $20\ \mathrm{dB}$, that the higher degree kernels, here $\kappa_{i,3,3}$, start to have a noticeable impact near the saturation point. Finally, it is worth noting that the simulations of the system model confirm with high accuracy the theoretical formula. 
\addtocounter{footnote}{-1}
%
%
\footnotetext{Kernels are simulated as partial sums where $k_i\in \{-L,-L+1,\hdots,L\}$.}
%
\subsection{Extension to Multicarrier Scenario}
%
The extension to more than one carrier is now considered ($N_T>1$). 
Two new phenomena must be taken into consideration: 
the power imbalance between the carriers\footnote{In this context, $P$ designates the total signal input power.} and the ACI caused by the carriers intermodulation products. 
On the one side, the useful part of the carrier power is dependent on the split of power between the carriers; it is denoted $C_{\mathrm{ISI},\{i\}}$, where $i$ represents the carrier index under consideration. This translates in an additional factor -- function of the power split -- arising in $c_1$, $c_2$ and $c_3$, which will not be the focus point of this paper. 
On the other side, an analytical expression of the ACI has been derived in \cite{Beidas2011IMD}. The variance of this expression is considered to compute $I_{\mathrm{ACI},\{i,\nu\}}$, where $\nu$ refers to the index of a carriers combination based on a list of possible combinations. 
In the frame of this paper, the ACI of order 1 and 3 are derived. 
%
For the sake of simplicity, the cross-terms between $I_{\mathrm{ISI}}$ and $I_{\mathrm{ACI},\{i,\nu\}}$ will be neglected along with the constructive terms contained in the ACI. 
%
%
Finally, the final CIR taking into account ISI and ACI is expressed as:
%
\begin{equation} \label{CIR_Total}
\left(\frac{C}{I}\right)_{\{i,\nu\}}\approx\left(\frac{C_{\mathrm{ISI},\{i\}}}{I_{\mathrm{ISI}}+I_{\mathrm{ACI},\{i,\nu\}}}\right)\ .
\end{equation}
Thus, by taking into account the main useful and interfering contributions, this expression will allow low-complexity CIR calculations.

%
\section{Dynamic Carrier Allocation}
\subsection{Scenario}
In this section, a carrier allocation scenario is considered to demonstrate the practicality of the proposed formula \eqref{CIR_Total}. A multicarrier signal composed of $N_c=3$ carriers $\{\mathcal{C}_i\}_{1\leq i\leq N_c}$ is considered at the HPA input. The carriers characteristics are listed out in Table~\ref{tabCarrier}. 
Since the signal power at the HPA input is $P$, a carrier with power split of 1/4 means that the input carrier power is $P/4$. The wideband OMUX depicted in Fig.~\ref{fig:DVBS2xCharac} is considered as post-HPA linear distortions characteristic. 
Since practical multicarrier systems rely to some extent on HPA linearization or predistortion techniques \cite{Beidas2016DPD} with input back-off closer to saturation, an equivalent input power of $P_{\mathrm{dB}}=-15\ \mathrm{dB}$ is chosen to translate these compensation mechanisms. This will also help to emphasize the contribution of each type of interference. 
%
\begin{table}[b!]
\vspace{-0.6cm}
\caption{Parameters for Carrier Allocation Scenario}
\vspace{-0.3cm}
\begin{center}
\begin{tabular}{|c|c|c|c|c|}
\hline
\textbf{Carrier}&\multicolumn{4}{c|}{\textbf{Characteristics}} \\
\cline{2-5} 
\textbf{ } & \textbf{$R_{s,i}\ \mathrm{(Mbauds)}$}& \textbf{$\alpha_{i}$}& \textbf{$B_{c,i}\ \mathrm{(MHz)}$}& Power split \\
\hline
$\mathcal{C}_1$ & 120.48 & 0.3 & 156.62 & 1/4 \\ \hline 
$\mathcal{C}_2$ & 120.48 & 0.2 & 144.58 & 3/8 \\ \hline 
$\mathcal{C}_3$ & 180.72 & 0.1 & 198.79 & 3/8 \\ \hline 
\end{tabular}
\label{tabCarrier}
\end{center}
\vspace{-0.2cm}
\end{table}
\subsection{Optimization Problem Formulation}
The center frequency of carrier $\mathcal{C}_i$ corresponding to the $\nu^{\mathrm{th}}$-possible carrier combination is denoted $f_{0,i,\nu}$. 
We define the set of all possible carrier placements $\mathcal{F}_{0}$ such that any $N_c-$tuple $\{f_{0,i,\nu}\}_{1\leq i\leq N_c}$ belonging to this set is a valid carrier allocation combination, where $\nu$ stands for the index of the combination. 
For clarity purposes, the solution space of the carrier allocation problem has been reduced to a discrete set with cardinality $|\mathcal{F}_{0}|=N_c!=6$ possibilities.
%

Although different optimization approaches can be considered, this paper will focus on a user fairness approach, which aims at maximizing the sum capacity. As described in \cite{Goodman}, this can be reduced to a max-min CIR optimization. 
The optimization problem can then be formulated as follows:
%
%
\begin{align} \label{Problem}
	&\mathcal{P}:\quad\underset{1\leq \nu\leq |\mathcal{F}_{0}|}{\mathrm{max}}\left\lbrace\min\limits_{\substack{1\leq i\leq N_c\\ f_{0,i,\nu} \in \mathcal{F}_{0}}}\left\lbrace\left(\frac{C}{I}\right)_{\mathclap{\ \ \ \{i,\nu\}}}\ (f_{0,i,\nu})\right\rbrace\right\rbrace
\end{align}
%
%
where the goal is to maximize, by carrier allocation, the minimum CIR between the carriers given the set of possible carrier placements. 
The key enabler in making carrier allocation dynamic is the computation speed of the CIR. This analysis has been performed in \cite{Colin} in the context of linear distortions, where numerical integration and slope-based distortions approximation was proven to be a good balanced between accuracy and complexity. Therefore, this computation method has been extended here to nonlinear distortions effects. 
%
%
%
%
%
\begin{figure}[t!]
	\centering 
	\scriptsize 
	\setlength\fheight{0.55\columnwidth} 
	\setlength\fwidth{0.55\columnwidth}
	\input{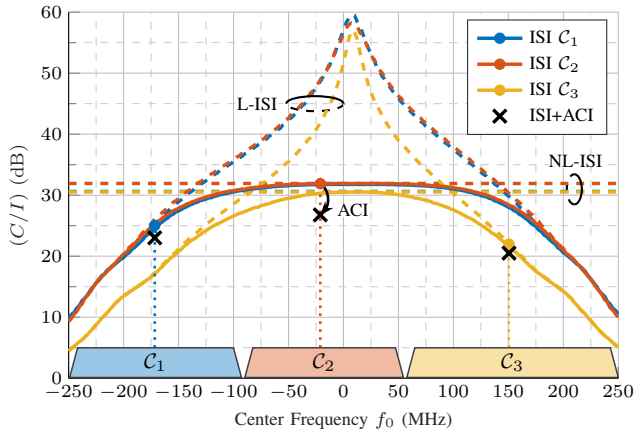}
	\vspace{-0.15cm} 
	\caption{Dynamic carrier allocation scenario. Multicarrier signal power at HPA input: $P_{\mathrm{dB}}=-15\ \mathrm{dB}$.}
	\label{fig:DCA}
	\vspace{-0.35cm} 
\end{figure}
%
%
\subsection{Results Discussion}
The carrier combination $\mathcal{C}_1 \mathcal{C}_2 \mathcal{C}_3$ is depicted in Fig.~\ref{fig:DCA}. As illustrated, the total CIR can be decomposed into types of interference: L-ISI, NL-ISI and ACI. 
In this scenario, L-ISI is critical on the edge of the filter and essentially depend on the symbol rate with small variations due to the roll-off. Indeed, $\mathcal{C}_3$ symbol rate is $1.5$ higher than the other carriers, therefore, it is subject to noticeable degradation in the order of $6\ \mathrm{dB}$. 
On the other hand, the NL-ISI is critical in the center of the filter and essentially rely on the signal input power $P$ with small variations due to the power split between the carriers and the roll-off. As such, the carriers $\mathcal{C}_2$ and $\mathcal{C}_3$ have a higher input power, and consequently, are getting more gain due to the power robbing effect. However, lower roll-offs degrade more the CIR, which counterbalances the power robbing effect to some extent in this specific scenario. 

\begin{figure}[t!]
	%
	\begin{minipage}[b]{1.0\linewidth}
	  \centering
	  \centerline{\resizebox{9.1cm}{!}{
%
%
\definecolor{mycolor1}{rgb}{0.00000,0.44700,0.74100}%
\definecolor{mycolor2}{rgb}{0.85000,0.32500,0.09800}%
\definecolor{mycolor3}{rgb}{0.92900,0.69400,0.12500}%
\definecolor{mycolor4}{rgb}{0.49400,0.18400,0.55600}%
%
\begin{tikzpicture}

\begin{axis}[%
width=4.71in,
height=0.9in,
at={(0.758in,0.509in)},
scale only axis,
xmin=1,
xmax=10,
xtick={1,2,...,10},
xlabel style={font=\large},
xlabel={\# CIR computations},
ymode=log,
ymin=0.0001,
ymax=10000,
ytick={0.0001,0.001,0.01,0.1,1,10,100,1000,10000},
yticklabels={{ },{$10^{-3}$ sec},{ },{ },{$10^{0}$\ \ sec},{ },{ },{$10^{3}$\ \ sec},{ }},
ylabel style={font=\large},
ylabel={Running Time},
axis background/.style={fill=white},
axis x line*=bottom,
axis y line*=left,
xmajorgrids,
ymajorgrids,
]
\addplot [color=mycolor1, mark=asterisk, line width=1.1pt, mark size=4pt, mark options={solid, mycolor1}]
  table[row sep=crcr]{%
1	16.2839815277778\\
2	32.5679630555556\\
3	48.8519445833333\\
4	65.1359261111111\\
5	81.4199076388889\\
6	97.7038891666667\\
7	113.987870694444\\
8	130.271852222222\\
9	146.55583375\\
10	162.839815277778\\
};

\addplot [color=mycolor2, mark=o, line width=1.1pt, mark size=3pt, mark options={solid, mycolor2}]
  table[row sep=crcr]{%
1	0.0262425772357723\\
2	0.0524851544715447\\
3	0.078727731707317\\
4	0.104970308943089\\
5	0.131212886178862\\
6	0.157455463414634\\
7	0.183698040650406\\
8	0.209940617886179\\
9	0.236183195121951\\
10	0.262425772357723\\
};

\addplot [color=mycolor3, mark=square, line width=1.1pt, mark size=3pt, mark options={solid, mycolor3}]
  table[row sep=crcr]{%
1	0.000370373983739836\\
2	0.000740747967479673\\
3	0.00111112195121951\\
4	0.00148149593495935\\
5	0.00185186991869919\\
6	0.00222224390243902\\
7	0.00259261788617885\\
8	0.00296299186991869\\
9	0.00333336585365853\\
10	0.00370373983739836\\
};

\end{axis}
\draw[draw,color=mycolor3,fill=white,opacity=1.0, line width=1.1pt] (12.6,1.75) circle (0.35)  node [] {{\color{black}TH2}}; 
\draw[draw,color=mycolor2,fill=white,opacity=1.0, line width=1.1pt] (11.3,2.25) circle (0.35)  node [] {{\color{black}TH1}}; 
\draw[draw,color=mycolor1,fill=white,opacity=1.0, line width=1.1pt] (9.9,3.0) circle (0.35)  node [] {{\color{black}SIM}}; 
\end{tikzpicture}%
}}
	\end{minipage}
	\vspace{-0.8cm}
	\caption{Comparison of estimated running time per method. Measurements were performed using a contemporary server processor clocked with $3.5\ \mathrm{GHz}$.} 
	\label{fig:Complexity}
	\vspace{-0.45cm}
\end{figure}
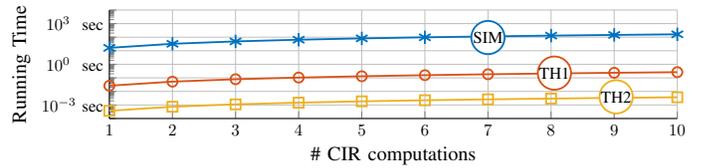

By performing an exhaustive search, the best carrier allocation is $\mathcal{C}_2 \mathcal{C}_3 \mathcal{C}_1$ with a minimum CIR of $23.64\ \mathrm{dB}$, whereas the worst combination is $\mathcal{C}_3 \mathcal{C}_2 \mathcal{C}_1$ with a minimum CIR of $19.77\ \mathrm{dB}$. This means that even with a low number of carriers it is possible to reach an allocation gain of $3.87\ \mathrm{dB}$. 
%
The order of magnitude of the complexity per CIR computation is highlighted in Fig.~\ref{fig:Complexity}. 
The curves denoted ``SIM'', ``TH1'' and ``TH2'' represent the CIR computation by conventional transmission chain simulation, the theoretical CIR formula without pre-computation, and the theoretical CIR formula with pre-computed kernels, respectively.  
Indeed, the proposed formula provides a significant reduction in the implementation speed (by a factor $6\cdot 10^2$ between ``SIM'' and ``TH1''). 
Since the ISI are independent from the carriers combination, the majority of the kernels can be computed beforehand when knowing the gain and group delay distortions, carriers symbol rate, roll-off, and power distribution. 
This provides further complexity reduction (by a factor $7\cdot 10^1$ between ``TH1'' and ``TH2''). 
In this scenario, the minimum, mean and maximum accuracy of the proposed theoretical formula with respect to the simulations are $97.38\%$, $98.74\%$, and $99.98\%$, respectively. The CIR formula leads to the same optimal allocation, which emphasizes the superiority of the proposed approach.
%
%

%
\section{Conclusion}
This paper provided a novel analytical formula, which characterizes the linear and nonlinear distortions in future UHTS systems. 
The behavior of different types of interference entailed by these distortions and their mutual interaction have been analyzed, bringing valuable insights. 
The principle of DCA has been highlighted in a realistic scenario, where an allocation gain of $3.87\ \mathrm{dB}$ can be achieved even with a limited number of carriers. 
Moreover, the proposed formula leads to a significant increase in implementation speed with respect to the conventional transmission chain simulations. 
These results are key to pave the way in considering near real-time DCA optimization for further capacity gains in the next-generation of satellite systems.
Finally, future research works will focus on more advanced algorithms able to solve the optimization problem within a shorter execution time. 

%


\end{document}